# Interactive framework for techno-economic meta-analysis of marine hydrogen transportation

Lucas Hanssens[1], Maarten Houlleberghs[1], Karel Van Acker,[2,3,4], Johan A. Martens[1,5] and Eric Breynaert[1,5,*]

## Abstract

As a result of the wide span of uncertainties and input variables considered in recent peer-reviewed studies on hydrogen supply chains, consensus and generalized insight are hard to derive. This work presents a meta-analysis model, building upon a growing database of literature cost estimations and enabling dynamic investigation and comparison of hydrogen transportation chains under varying input scenarios. Also built into the framework is the evaluation of the impact of costing methodology and shipping fuel type. A convenient online interface allows to leverage increasing amounts of underlying cost data and assists users in gaining insight into generalized cost dynamics attributes. The method and toolbox are demonstrated via in-depth analysis of compressed hydrogen ($GH_2$), liquefied hydrogen ($LH_2$), dibenzyltoluene (LOHC) and ammonia as hydrogen carriers. Confined hydrogen clathrate hydrates are used as an example to demonstrate how one can deal with lower-TRL candidate technologies.

## 1 Introduction

Avoiding drastic climate change, induced by excessive emission of greenhouse gases, requires a dramatic turnaround. In the European Green Deal, the EU has pledged to guide its economy and society into climate-neutrality by 2050 [1]. These ambitious plans demand swift adoption of renewable energy strategies, also in geographic regions with limited potential for domestic production of renewable energy [2,3]. Currently these regions rely on the import of fossil energy for industrial production and power supply, but multiple options for decarbonization of key industrial production processes exist. Import of green hydrogen could for example replace import of fossil fuels [4]. On the flip side, there is the possibility of re-localization of energy-intensive production processes to foreign regions with large renewable energy potential (an effect known as "renewables pull") [5,6]. The extent to which (part of) the value chain will be relocalized, critically depends on a complex interplay of the perspectives of private investors, policy makers and society on a given product [7]. The transportation cost for renewable energy, such as green hydrogen, clearly is a critical factor in this consideration. Several marine transportation technologies can be considered for rapid scale-up of green hydrogen imports, but the optimal choice will depend on specific use cases. This renders the techno-economic comparison between different scenarios far from straightforward and calls for high-quality and nuanced techno-economic assessment of hydrogen transportation.

[1] *Center for Surface chemistry and Catalysis, KU Leuven, Celestijnenlaan 200F – box 2461, 3001 Heverlee, Belgium*

[2] *Department of Materials Engineering, KU Leuven, Kasteelpark Arenberg 44, 3001, Leuven, Belgium*

[3] *Flanders Make, Corelab, VCCM, Belgium*

[4] *Center for Economics and Corporate Sustainability (CEDON), KU Leuven, Warmoesberg 26, BE-1000, Brussels, Belgium*

[5] *X-ray/NMR platform for Convergence Research (NMRCoRe), KU Leuven, Celestijnenlaan 200F – box 2461, 3001 Heverlee, Belgium*

In recent years, literature has seen a sharp surge in techno-economic studies on hydrogen supply chains. Especially for marine supply chains, the large capital expenditures (CapEx) and long lifetimes of the infrastructure require strong commitment to a certain technology. Table 1 shows an extensive overview of recent peer-reviewed studies investigating the potential of a wide range of hydrogen storage technologies under different marine supply chain assumptions. To make case-specific recommendations, most studies focus on real-world cases. Investigating costs along well-defined trading routes holds potential to highly impact policymakers looking to establish such routes [8–11]. Comparing the case studies in Table 1, a very large span of supply chain variables emerges. This not only has led to a diverse range of case-specific technology optima (indicated in bold in Table 1), the strongly differing supply chain assumptions render benchmarking across case studies extremely difficult [12]. Clearly, there is no "silver bullet" in hydrogen supply chain literature. Even studies considering similar routes and using similar technologies tend to draw contradictive conclusions, due to the large variability of input variables (e.g. Ammonia cracking CapEx estimations can vary up to an order of magnitude for infrastructure of similar capacity). This lack of consistency is confusing and lowers the confidence of policymakers and energy system researchers towards the outcomes of such studies.

A major barrier to consensus is the absence of standardized frameworks that allow for dynamic, comparative analysis across a wide range of input assumptions and costing methodologies. The studies in Table 1 mostly rely on different, case-specific models and often also apply different techno-economic assessment strategies. In most cases, they are either based on equivalent annual cost (EAC) or on net present value (NPV) methods. None of these studies systematically address the influence of the method used on conclusions of the study. As will be shown in the present study, the choice of economic assessment strategy can in itself have a large impact on the conclusions indicated by the modelling. Also key assumptions such as discount rate, asset lifetime, and energy cost vary widely between studies. Often this is insufficiently disclosed.

Broader, interactive, energy system modelling tools, such as HOMER, TIMES, OSeMOSYS or PLEXOS, offer system-level insights into hydrogen flows but lack the process-level resolution to differentiate between marine carrier technology configurations and these models include limited specific literature data [13–16]. They are also not easily and freely accessible, or quickly adaptable for rapid sensitivity or scenario analysis focused only on the hydrogen supply chains.

To address these limitations, the present work introduces a flexible and publicly accessible meta-analysis modelling framework that builds on a growing database of specialized techno-economic literature. The tool enables transparent comparison of hydrogen transport technologies under varying input conditions, including transport distance, energy price, infrastructure assumptions, fuel type, and costing methodology. By allowing users to explore the impact of such variables interactively, the model facilitates both high-level scenario analysis and granular sensitivity exploration. In addition to established carriers such as $GH_2$, $LH_2$, ammonia, and LOHC, the framework supports the inclusion of low-TRL technologies, broadening its applicability to emerging options. Finally, our dashboard is designed to be intuitive. This provides experienced as well as less experienced readers the opportunity of gaining insight into cost dynamics. It makes our dashboard truly unique in its ability to engage researchers and policymakers that would otherwise not be able to explore their scenarios of interest, as most interested researchers, and especially policymakers, do not have access to licensed energy system modelling tools. To the best knowledge of the authors, no comparable open-access meta-analysis tools are available for a broad readership to get insight into the specific literature of techno-economic cost determination of hydrogen supply chains.

This work kicks off with an elucidation of the input categories causing a large part of the variability in cost outcomes. These input categories are critical parameters included in the tool input dashboard. After introducing the general structure of the model, the method and toolbox are demonstrated by providing an in-depth analysis of compressed hydrogen ($GH_2$), liquefied hydrogen ($LH_2$), dibenzyltoluene (LOHC) and ammonia. The choice for these technologies was made due to their prevalence in the present literature. Pipelines are not considered in this demonstration of the framework because of their limited application potential for trans-oceanic transport, but can be added to the framework in future work, providing improved results for short to medium distance routes. To demonstrate how one can deal with lower-TRL candidate technologies, confined hydrogen clathrates were added to the analysis. The discussion of the results focuses on the identification of technology-specific cost drivers and underlying trends in the cost dynamics of the different technologies. Finally, the capability of the model to research different scenarios is demonstrated with a selection of sensitivity and scenario analyses deemed interesting by the authors.

The data and analysis framework used in the study are available in the interactive online application '[Hydrogen Supply Chain Dashboard](#)', which can be accessed at 'https://script.google.com/macros/s/AKfycbwoSX4K5dUJmykwXWmtsKj_UKAF_z0pdDaBjuj9j4v-LSE6zgas3RPNsTnhx-TCMLVHcg/exec'. If the app does not open, using a private window can solve the issue. Further inquiries can be directed to the corresponding author.

**Table 1:** Overview of techno-economic analyses of hydrogen supply chains published starting from 2017. For comparative assessments, cost-optimal technologies are indicated in bold. EAC: Equivalent Annual Costing, LCOT: Levelized Cost of Transport, LCCA: Life-Cycle Cost Analysis.

| | Technologies considered | Fuel for transport | Method | Reference |
|---|---|---|---|---|
| Chapman *et al.* (2017) | $LH_2$ | Fossil | EAC | [8] |
| Niermann *et al.* (2019) | LOHC | Fossil | EAC | [17] |
| Heuser *et al.* (2019) | $LH_2$ | Fossil | EAC | [9] |
| Ishimoto *et al.* (2020) | **$LH_2$**, Ammonia | Renewable | $LCOT_{NPV}$ | [11] |
| Niermann *et al.* (2021) | $GH_2$, $LH_2$, **LOHC**, HVDC, Pipeline | Fossil | EAC | [18] |
| d' Amore Domenech *et al.* (2021) | $GH_2$, **$LH_2$**, HVDC, Pipeline | Renewable | LCCA | [19] |
| Gallardo *et al.* (2021) | $LH_2$, **Ammonia** | Fossil | Not clear | [20] |
| Raab *et al.* (2021) | $LH_2$, **LOHC** | Renewable | EAC | [21] |
| Brändle *et al.* (2021) | $LH_2$, **Pipeline** | Renewable | EAC | [22] |
| Roos (2021) | $LH_2$, LOHC, **Ammonia** | Fossil | EAC | [23] |
| Collis *et al.* (2022) | $GH_2$, $LH_2$, **LOHC**, **Ammonia**, **Pipeline** | Not clear | Not clear | [24] |
| Lee *et al.* (2022) | $LH_2$, **LOHC**, Ammonia | Fossil | EAC | [25] |
| Borsboom-Hanson *et al.* (2022) | $GH_2$, **$LH_2$**, **Pipeline** | Fossil | EAC | [26] |
| Johnston *et al.* (2022) | $LH_2$, LOHC, **Ammonia**, S-LNG | Fossil & Renewable | EAC | [10] |
| Dickson *et al.* (2022) | $LH_2$, LOHC, **Ammonia**, S-LNG | Fossil | DCF | [27] |
| Perey and Mulder (2022) | $LH_2$, **Ammonia**, **Pipeline** | Fossil | $LCOT_{NPV}$ | [28] |
| Cui *et al.* (2023) | **Ammonia**, Methanol | Not clear | $LCOT_{NPV}$ | [29] |
| d' Amore Domenech *et al.* (2023) | **$GH_2$**, **$LH_2$**, **Pipeline** | Renewable | LCCA | [12] |
| Neumann *et al.* (2023) | $LH_2$, **Fe** | Fossil | EAC | [30] |
| Hampp *et al.* (2023) | $LH_2$, LOHC, Ammonia, S-LNG, FTF, $CH_4$ (g), **Pipeline** | Renewable | EAC | [31] |
| Villalba-Herreros *et al.* (2023) | **$LH_2$**, Ammonia | Renewable | EAC | [32] |
| Kim *et al.* (2024) | $LH_2$ | Renewable | EAC | [33] |

**Hydrogen supply chain input categories**

Techno-economic analyses on hydrogen supply chains typically define case studies using a set of parameters including the investor's perspective, the modelling scope and supply chain scale and distance. While these inputs can vary widely between studies, they are uniquely defined for each specific case. The most common hydrogen supply chain input variables can however be categorized into three main groups: 'Case Specification', 'Location' and 'Technology' (Figure 1). Even though case-related assumptions cannot always be aligned, the consistency in the underlying methodology should be increased.

'Case Specification' - To improve comparability between studies, this work identifies two important factors: (1) the fuel-type choice and (2) the sensitivity of the results towards different techno-economic costing methodologies. Modelling the impact of renewable fuels on supply chain costs for competing technologies offers an avenue to evaluate the resilience of the technology towards future fuel adaptations and increasingly stringent regulations governing supply chain emissions. As shown in Table 1, there is a large variability in the fuel type considered in current literature, but in most published cases the impact of fuel type on the analysis is barely discussed. Table 1 also highlights the diversity of cost assessment methods used in the consulted studies. As awareness on the impact of the different assessment methods is critical, this study provides an in-depth discussion on assessment methods, mathematically deriving specific conditions for equivalence of the most popular methods, while providing the opportunity to assess case-specific costing methodology impacts in the modelling dashboard.

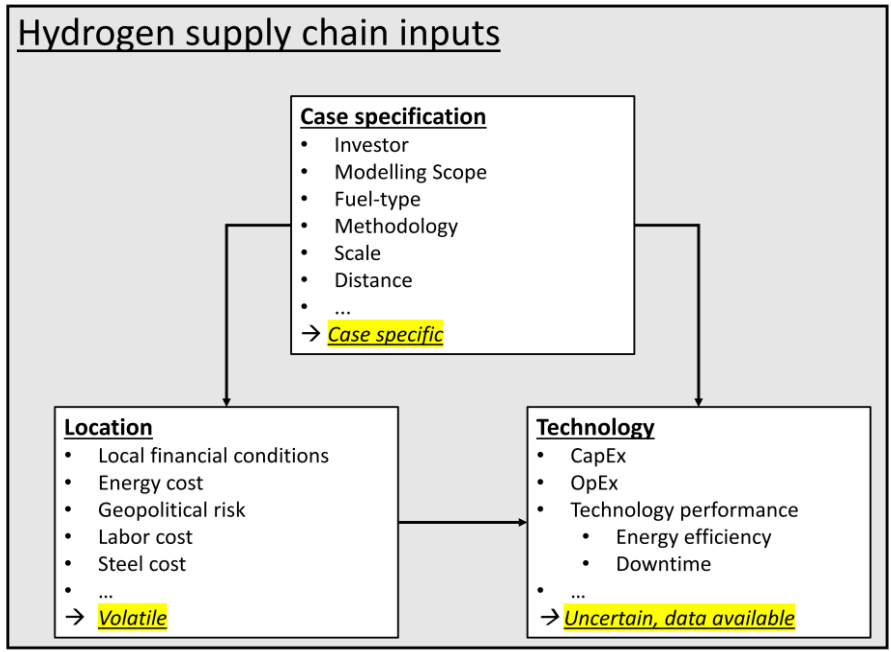

**Figure 1:** Diagram summarizing three types of input variables in hydrogen supply chain techno-economic studies.

'Location' - The location of infrastructure determines localized parameters including energy cost, climate and labour cost. These parameters tend to be very volatile in nature and can cause strong discrepancies in modelling outcomes [20].

'Technology' - Based on case specifications and location, specific choices are made for transportation and stationary infrastructure. As large-scale hydrogen supply chains are still in a conceptual stage, there is a high degree of uncertainty on how capital and operational costs for these technologies will evolve over the coming decades. Previously published studies often rely on a single process simulation or reference datapoint for estimating the capital and operational expenditures (CapEx & OpEx) of infrastructure that has yet to be developed at scale. To illustrate the heterogeneity in technology-related input data across different techno-economic studies on marine hydrogen supply chains, Figure 2 shows the distribution of (de)conditioning energy use and CapEx data extracted from a selection of published studies [8,11,19,20,22,25,27,29,34–47]. The presented data was normalized to 1 $GW_e$ and adjusted for inflation with respect to reference year 2022 (according to equation 7). These datapoints diverge strongly as authors have (1) varying degrees of confidence in the development of (de)conditioning infrastructure at GW-scale, (2) different supply chain boundary conditions (hydrogen temperature, pressure, purity, …), (3) different process modelling methodologies, including varying modelling scopes and detail.

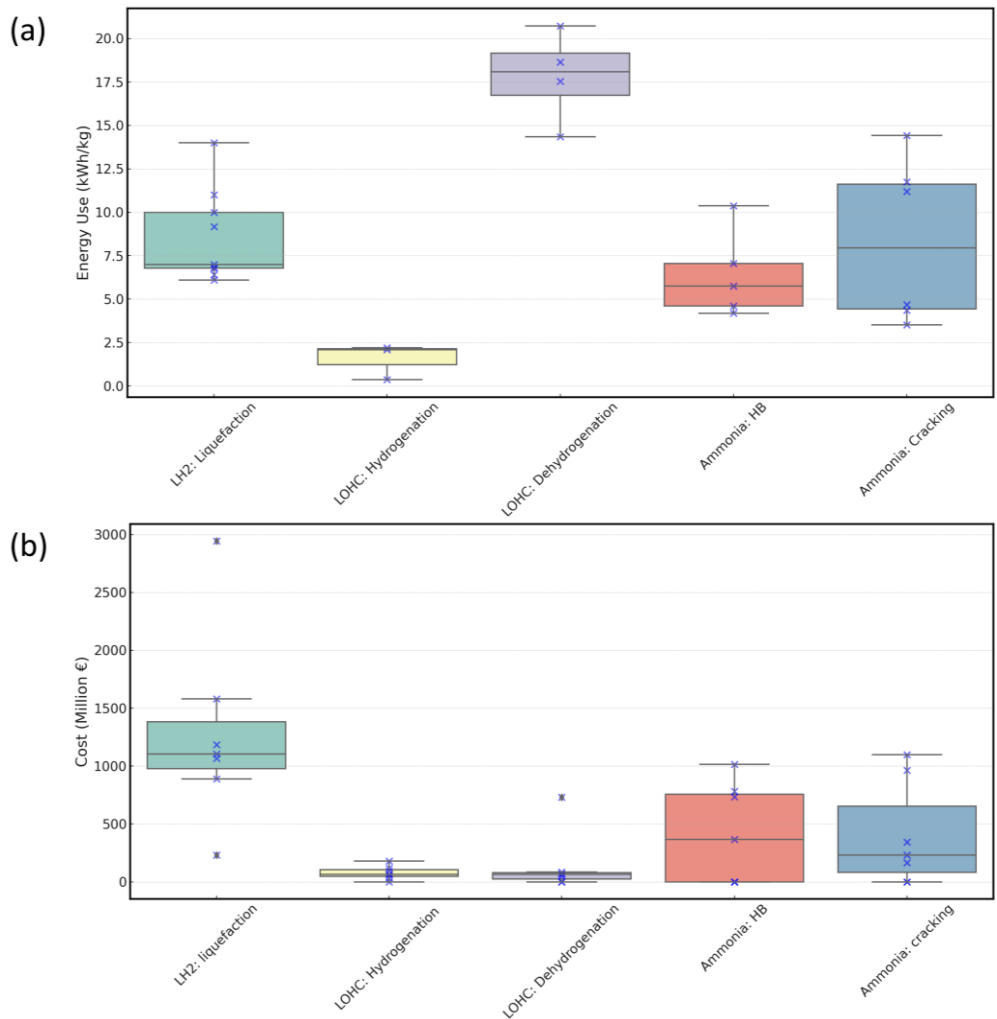

**Figure 2:** Boxplots illustrating the heterogeneity of input data on (a) (de)conditioning energy use (in kWh$_e$/kgH$_2$) and (b) CapEx for supply chain unit operations (normalized to 1 GW$_e$ and adjusted for inflation with respect to reference year 2022), used in a selection of techno-economic studies on hydrogen supply chains [8,11,19,20,22,25,27,29,34–47]. Specific information on the datapoints used can be found in the Supplementary Information (A1.3-A1.5). Scale-related discrepancies in the data were minimized by sourcing data from studies with modelling scales within a factor 5 of 1 GW$_e$.

## 2 Experimental procedures

The components of the marine transportation chain for the different technologies are schematically shown in Figure 3. The model calculates the transportation cost of a set amount of hydrogen delivered in gaseous form at 50 bar. As this work focuses on the transportation of the hydrogen, only hydrogen produced for compensation of hydrogen losses along the transportation chain are included in the cost assessment, to represent the transportation efficiency of the included technologies. Hydrogen losses occur through boil-off (LH$_2$) and also include hydrogen used for dehydrogenation heat (LOHC and NH$_3$) and propulsion. The modelling in this work starts from gaseous H$_2$ at low pressure. Then, through the application of the different conditioning technologies, the volumetric energy density of the hydrogen is increased to allow for efficient transportation. Conditioned hydrogen is stored in buffer reserves adapted to the time between the arrival of ships. An essential part of the transportation chain estimations concerns the scaling of ship fleets. The size of the fleet depends on the quantity of hydrogen that needs to be delivered and the number of deliveries per day for each ship (typically less than one), which depends on the distance from production to delivery site, according to Eq. 1:

$$\#\text{transport ships} = \left\lceil \frac{\frac{\text{daily production [kg H}_2\text{]}}{\text{capacity transport ship [kg H}_2\text{]}}}{\frac{\#\text{deliveries}}{\text{day} \cdot \text{ship}}} \right\rceil \quad \text{Eq.1}$$

Inherently, vehicle supply chains scale imperfectly, as no fractional vehicles can be employed. This attribute is represented by the ceiling operator in Eq.1. Optimal scaling points are achieved when all vehicles leaving the production facility are loaded at maximal capacity. This is a situation in which the buffer storage capacity is also optimally sized.

In the base case, used to demonstrate the model, the capacity of the marine supply case is set at 1 GW$_e$, corresponding to ca. 426 tonnes hydrogen per day. The input hydrogen pressure of the marine supply chain is assumed to be 5 bar. The output pressure is set at 50 bar, since it is assumed that at the receiving port, these large volumes of hydrogen will be loaded into a pipeline network [48]. Large ports often serve chemical clusters with interconnected hydrogen pipelines [49]. Further details on the build-up of the supply chains for the different technologies can be found in the Supplementary Information (A1.2-1.6).

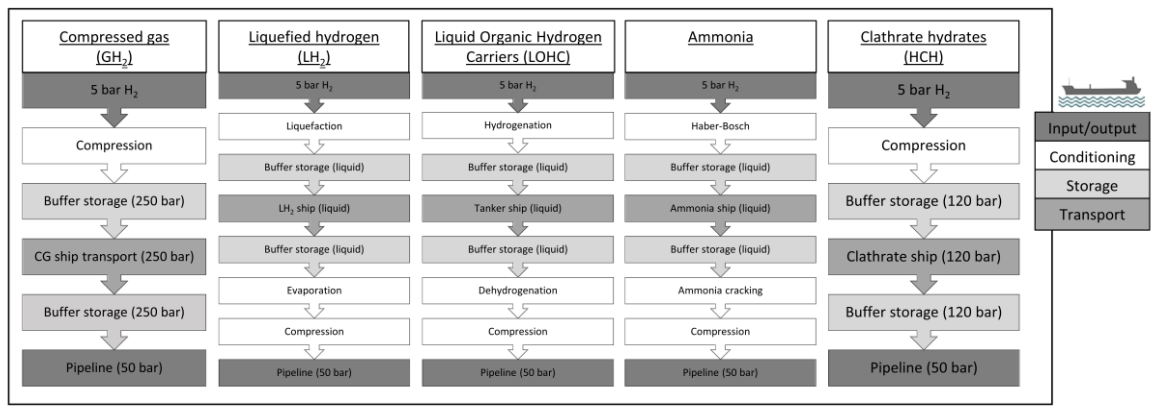

**Figure 3:** Modelling scope and components of the marine transportation supply chain.

## 2.1 Model structure

The modelling process is represented as a flowchart in Figure 4. A technology modelling back-end model (running in Google Sheets supported by automatization and solver protocols) is published on the web via a front-end application dashboard (Figure 5), optimized to structure user interactions with the back-end. The supply chain configuration and techno-economic assumptions are defined in the application dashboard. After submitting these inputs to the back-end technology modelling framework, all technologies are modelled separately following the same basic modelling outline (Figure 3). First, the model is initialized using data from the database and the application dashboard. This yields the supply chain infrastructure dimensioning. The toolbox offers two options for using the available data: all data can be averaged, thus representing a general cost trend, or alternatively, one can select and compare specific data. Subsequently, process energy flows are calculated based on reference input data and process modelling. Finally, the supply chain infrastructure and energy requirements are valorized in a discounted cashflow analysis. In the following sections, the outline of each of these modelling blocks is further discussed.

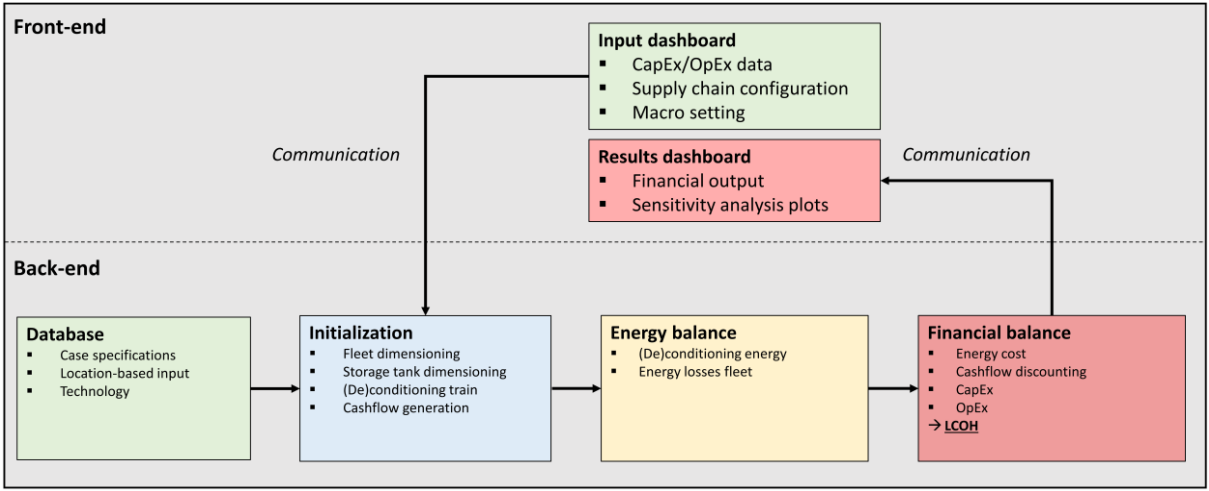

**Figure 4:** Modular structure of the modelling framework.

![Dashboard rendering]

**Figure 5:** Rendering of the supply chain web-app dashboard.

### 2.1.1 Initialization Module

#### 2.1.1.1 Inputs from database and applications dashboard

The initialization module receives input, both from the back-end database and from the front-end application dashboard. The back-end database delivers input on the hydrogen transportation technologies considered and default costs associated with these technologies. These inputs are combined with the inputs provided by the application dashboard, including the supply chain configuration, the macroeconomic setting, as well as modifications to the default inputs retrieved from the back-end database. In what follows, the default database configurations and their associated costs are outlined.

*2.1.1.2   Hydrogen transportation technologies*

Table 1 shows an overview of the technologies collected from literature. The database contains information on well-established technologies including compressed hydrogen, liquefied hydrogen, LOHC (Dibenzyl toluene), and ammonia, as well as on one emerging technology: confined hydrogen clathrate hydrates. The option to define low Technology Readiness Level (TRL) technologies such as HCH and compare their performance to established technologies can yield valuable insights and is an asset of the presented methodology.

An important disadvantage of $H_2$ as molecular energy carrier is its low volumetric energy density at atmospheric pressure. To compensate this, it is mostly stored and transported at high pressure or in liquefied state. Both technologies are mature at small scale and have been widely implemented across a broad range of applications. Technically, compressed hydrogen storage ($GH_2$) is the least complex $H_2$ storage technology. Five types of storage vessels are available: Type I to V. Type I vessels are typically employed for stationary storage at pressures of up to 500 bar. They are mostly constructed in austenitic stainless steel alloys (e.g. AISI 304 and AISI 316). While these alloys are best fit to withstand hydrogen effects at ambient temperatures [50,51], their high specific density yields heavy vessels with limited application potential in transport. Lightweight vessels can be constructed using composite materials and are more expensive, but they are more suitable for transport of $H_2$. For Type II vessels, the load is distributed between the metal tank and a fiber wrap. Type III and type IV vessels consist of a thin liner, respectively in metal or high-density polyethylene, wrapped in a carbon fiber reinforced resin based shell, which carries the load [52,53]. Type V vessels consist entirely of composite material.

Many disadvantages have been reported regarding the use of compressed or liquefied hydrogen as energy vectors in future sustainable energy distribution scenarios [18,19,36,54]. Major drawbacks for compressed $H_2$ gas are the high energy penalty for compression, the need for heavy and/or expensive pressure vessels, and the specific safety precautions and regulations that need to be taken into account. Compared to gaseous $H_2$ storage, the gravimetric energy density of liquified $H_2$ is higher. Its extremely low boiling point (20 K) however limits the options for storage and transport to costly vacuum-insulated vessels. Storage and transportation of liquefied $H_2$ also comes with a low energy efficiency as result of its extremely low temperature. Today, hydrogen liquefaction typically consumes between 30 and 40% of the chemical energy present in the liquid (on a lower heating value basis). Boil-off during transport leads to loss of $H_2$ in gaseous form [18], further decreasing the efficiency. In an attempt to improve the economics of $H_2$ storage and transport, alternative technologies for $H_2$ storage and transportation are in high demand. Several are being developed and investigated, each with their own strengths and challenges.

One of the alternative technologies makes use of liquid organic hydrogen carriers (LOHC) [2,18,54,55]. LOHCs are hydrocarbon molecules (often aromatic), occuring as liquids at ambient pressure and temperature, which can catalytically be hydrogenated to store chemically bonded hydrogen. $H_2$ is recovered by catalytic dehydrogenation of the organic molecules. After returning the dehydrogenated carrier molecules to the hydrogenation site, the cycle is closed. Methanol, benzyltoluene and dibenzyltoluene (DBT), N-ethylcarbazole and 1,2-dihydro-1,2-azaborineare are typical LOHC candidates [17,55]. As these compounds exhibit physicochemical characteristics similar to diesel fuel, they can be transported using the existing and well-established infrastructure for distribution of oil derivatives. In this application, the heavier, aromatic molecules are typically preferred because of their low vapor pressure. This minimizes evaporative losses and facilitates purification of the released $H_2$ to fuel cell grade [56,57].

Another attractive hydrogen carrier is ammonia [41,58]. The gravimetric and volumetric capacity of hydrogen stored in liquid ammonia surpasses that of liquid hydrogen [41,59]. Ammonia not only is already being produced on an industrial scale using an electrified Haber-Bosch process [58], it is transported around the globe in large volumes for fertilizer production and chemical applications [60]. This implies large parts of the supply chain are already available. While these advantages make ammonia an attractive candidate for established marine transport businesses to engage in hydrogen transportation [61,62], ammonia technology also comes with severe downsides. Ammonia not only is extremely toxic and environmentally unfriendly, releasing its $H_2$ content requires an energy-intensive cracking step. Ammonia cracking can be done using fire-heated cracking furnaces, much like steam methane reforming, but more effective (catalytic) cracking methods are actively being researched [45]. To the best of our knowledge, there are currently no large scale ammonia cracking facilities operational. Significant costs for purification are anticipated since $H_2$ applications typically have very low tolerance for ammonia contamination of the hydrogen feed [63]. Uncontrolled release and subsequent precipitation of atmospheric ammonia in natural reserves and large water bodies is problematic. Its toxicity further limits its application potential, especially when transport on public roads in densely populated areas is required [41,64,65]. This implies strict safety measures will have to be put in place to unlock the potential of ammonia as a widespread energy carrier.

Finally, $H_2$ ad- or absorbing materials such as activated carbon, metal organic frameworks and metal hydrides could play a role in future hydrogen supply chains [35]. Hydrogen storage in clathrate hydrates is another appealing option [66,67]. Clathrate hydrates are solid ordered water structures encapsulating guest molecules in molecular cages, at high storage densities. Methane filled clathrate hydrates are abundant in nature, occur in deep-sea sediments and permafrost [68,69], and are investigated as a potential alternative to LNG in marine natural gas supply chains [70–74]. Hydrogen clathrates do not occur in nature, but can be made synthetically [66,67,75,76]. The hydrogen storage capacity of hydrogen clathrate hydrates (HCH) depends on its crystallographic structure. Storage capacities of 3.2 wt% and 5.2 wt%, have been reported for structure I (sI) and structure II (sII) $H_2$ hydrates, and theoretically, clathrate structures could store up to 10 wt% $H_2$ [76]. Using confined HCH, this technology could even be operated similar to an adsorption process. To the best of our knowledge, HCH technology has not yet been considered in the existing techno-economic literature, rendering it a valuable addition. Progress is being made in the understanding of the molecular aspects of synthetic clathrate formation [69]. Crystallization of bulk clathrate hydrates is particularly challenging and requires extreme pressures of several thousand bar or deep cooling. The use of tetrahydrofuran (THF) as clathrate crystallization promotor has resulted in a dramatic decrease in pressure and temperature requirements for the formation of sII hydrogen clathrate. Lee *et al.* achieved 4 wt% hydrogen storage density in sII clathrate hydrates stabilized using 0.2 mol% THF at 120 bar and 270 K [77]. Similar to THF, organic crystallization aids and hydrophobic surfaces can alleviate pressure and temperature requirements and improve both kinetics and thermodynamics of clathrate formation [78–84]. For the HCH supply chains used in this work, we used the results and conditions reported by Lee *et al.* and assumed that the promoting effect of THF can be matched by hydrophobic materials.

*2.1.1.3   Cost data*

2.1.1.3.1   CapEx

Supply chain costs are broken down into CapEx and OpEx. In the available literature, typically, CapEx data is obtained either from process simulations or reference data. Chemical processes, however, can be very complex and cost outcomes are heavily dependent on process configuration and assumptions. As illustrated in the introduction, cost estimates in literature for emerging technologies, involved in large-scale hydrogen supply chains, can vary substantially. Relying on a single process simulation outcome for the complex chemical processes involved in hydrogen supply chains thus introduces a high degree of uncertainty into the cost outcomes. Therefore, to increase the robustness of our analysis, we incorporated an average of at least 3 reference data sources in our model as a more nuanced proxy for capital intensive infrastructure investments. In order to improve data compatibility, reference data was only selected when the reference modelling scale was within a factor 5 of the standard modelling scale of 1 GW used in this work. A detailed description of included CapEx data can be found in A1.2-1.7. Indirect capital costs are assumed to be incorporated in the reference cost data, although we were unable to verify this for some datapoints due to their limited documentation.

To date, no marine hydrogen supply chains have been established at scale. As a result, conceptual hydrogen carriers were compiled for the different hydrogen transportation technologies. Ship design details are summarized in Table 2 and are more elaborately available in section A1.7. The conceptual $GH_2$ and HCH ship design is shown in Figure 6 and is based on available compressed natural gas carrier designs by Global Energy Ventures© [85]. Given the larger amount of available reference data, $LH_2$, LOHC and ammonia carrier designs were based on reference data. For each carrier, cost estimations are included for adaptations towards a renewably powered drivetrain.

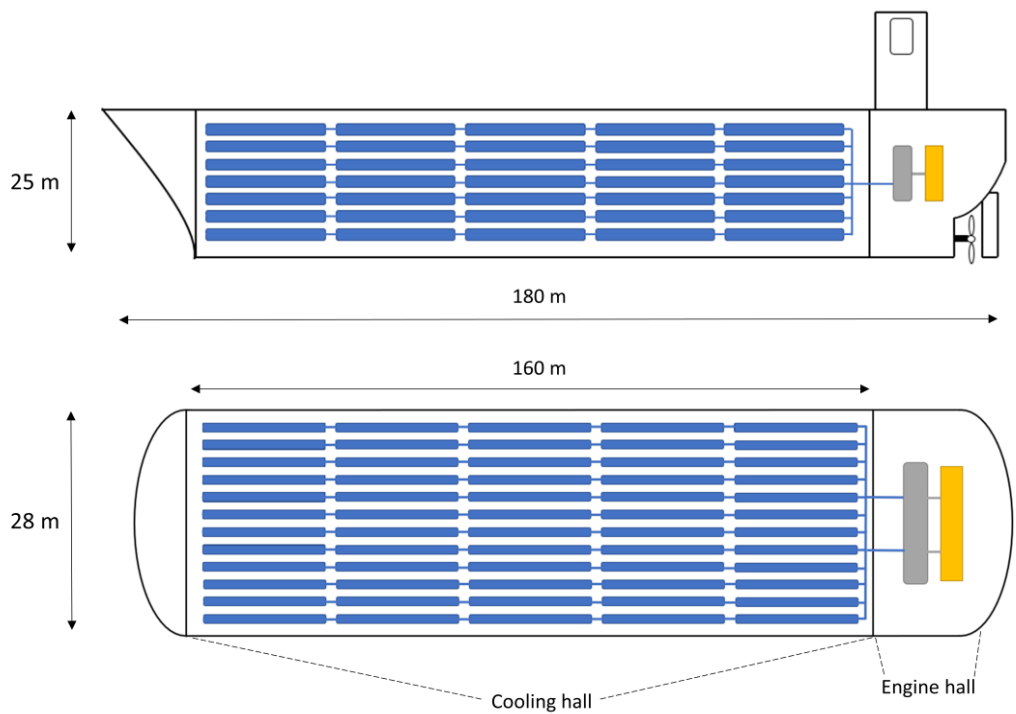

**Figure 6:** Conceptual HCH and $GH_2$ ship design. The standard design for both ships involves a hall with stacks of interconnected pressure vessels (for the HCH ship the hull is insulated).

**Table 2:** Marine hydrogen ship cost data and assumptions. Carrier molecule costs are included in LOHC ship cost estimations. Further detail is provided in section A1.7.

|  | $GH_2$ | $LH_2$ | LOHC | Ammonia | HCH |
|---|---|---|---|---|---|
| Capacity ship (ton H2) | 300.54 | 9828.31 | 2400.00 | 7040.00 | 1283.00 |
| Cost ship (M€) | 251 | / | 187 | 65 | 197 |
| Cost renewable ship (M€) * | 271 | 397 | 207 | 80 | 210 |
| Lifetime ship (years) | | | 20 | | |
| (Un)loading time (h) | 36 | 48 | 48 | 48 | 60 |

*: Including hydrogen storage agent (DBT for LOHC and porous material for HCH)

#### 2.1.1.3.2 OpEx

Averaged reference data is used to estimate the energy requirements of the more chemically complex process chains ($LH_2$, LOHC and ammonia), while $GH_2$ and HCH chain energy use is modelled on a unit operation level using empirical relations for compressor energy use and Carnot efficiency calculations for cooling performance (A1.1.1 and A1.1.2). General OpEx assumptions of this study are detailed in Table 3. In the default setting, supply chains are assumed to be powered by fossil fuels. IFO 380 bunker fuel prices from the first quarter of 2021 were used. Marine hydrogen supply chains will be deployed to transport energy from regions with abundant renewable energy resources, to regions with a deficit in renewable energy. For the low-cost producer regions, we assumed an energy cost of 0.021 €/kWh, corresponding to the low-end figure of current LCOE estimates for onshore wind energy in the Lazard ltd. LCOE report [86]. The electricity cost for the consumer region is based on the average European non-household electricity price reported by Eurostat [87]. 2021 averages were incorporated to avoid using the disproportionately inflated 2022 energy prices.

**Table 3:** Overview of the specific operational expense assumptions.

| | |
|---|---|
| Labor costs [88] | 19.8 €/h |
| Electrolysis efficiency | 52.59 kWh/kgH$_2$ |
| Maintenance costs [45] | 5.0 % of capital investment |
| Insurance costs [45] | 2.0 % of capital investment |
| Electricity cost producer region [86,87] | 0.021 €/kWh |
| Electricity cost consumer region [87] | 0.130 €/kWh |
| Fossil fuel cost [89,90] | 309.73 €/tonne |
| Fuel consumption [91] | 90 tonne/day |
| Working hours per day | 12 |
| Size crew [88] | 22 |
| Speed [91] | 35 km/h |

### 2.1.2 Financial balance

To make a direct comparison, the complex cost structure of each technology must be distilled into one appropriate performance indicator. Deriving a single cost figure from a supply chain, however, is an intricate task. Capital expenditures (CapEx) and operational expenditures (OpEx) occur scattered throughout the lifetime of the project. Consequently, the total cost figure is not merely a summation of costs, it must also consider concepts such as opportunity cost and time preference. In the case of hydrogen supply chains, the performance indicator of choice is often the levelized cost of hydrogen transport (LCOT), expressed in a certain currency, per mass or energy unit transported hydrogen. As in wholesale energy markets, prices are generally reported in €/MWh, since estimating the transport costs per MWh simplifies interpretation in the context of the energy market as a whole.

### 2.1.2.1 *Levelized cost of hydrogen transport (LCOT)*

Ideally, the LCOT can be compared directly to the levelized cost of energy (LCOE), which is a very impactful and widely implemented measure of energy cost [92,93]. Consequently, LCOT assessment methods are mostly analogous to the different LCOE assessment strategies present in literature. Currently, as Table 1 highlights, the predominant hydrogen-related LCOT approach is the equivalent annual cost strategy (EAC). EAC involves calculating the total yearly supply chain cost by summing annualized asset-specific CapEx contributions and yearly OpEx, akin to the LCOE as used by the National Renewable Energy Laboratory [92]. Some advantages of this method are its simple calculation, without the need for complicated data setup, the ability to compare costs for assets with different lifetimes and the ease of communicating a yearly cost figure [94]. The annuitization is done using a Capital Recovery Factor (CRF), for a specified depreciation period or over the complete useful lifetime. Importantly, using this method theoretically implies infinite project duration when the respective depreciation times of the considered appliances are different. In techno-economic assessments of a single asset, like LCOE calculation of renewable energy generators, the assessment period often coincides with the useful lifetime of the asset. However, the situation becomes more complicated in projects in which many assets are combined with different useful lifetimes. In that case the annualized CapEx of the different assets is combined in an overall system annuity (Equation 2 and 3). The respective annualized cost figures of the different assets represent a yearly cost considering their specified depreciation or lifetime. When the useful lifetimes of the assets in the system do not correspond, one cannot specify a certain project duration or assessment period without skewing the results. For infrastructure with a lifetime lower than that project duration, continuous replacement would be assumed, and the project duration would have to be a whole multiple of the asset's lifetime. For long-lived infrastructure costs are spread over a period that is either longer than the assessment period (when their useful lifetime is used), or over a period that is significantly shorter than their useful lifetime (if their depreciation time is matched with the project duration). This makes the EAC method in its standard form intrinsically awkward in dealing with strictly defined assessment periods and connected thereto, remaining (salvage) value of assets at the end of the project duration or assessment period. One should thus be careful applying the EAC method to multi-asset systems. To enable the inclusion of residual values of assets at the end of an assessment period, we included an adapted version of the EAC method (Eq. 4 and 5). In this version of the EAC method (referred to as EAC2), all CapEx contributions are annualized over the same assessment period. The CapEx figure to be annualized is determined based on how the asset's depreciation time ($t_d$) compares to the assessment period.

$$\text{EAC} \left[\frac{\text{€}}{\text{MWh}}\right] = \frac{\text{OpEx}_{\text{yearly}} + \sum_i \text{CapEx}_i \cdot \text{CRF}_i}{\text{E}_{\text{yearly}}} \quad \text{Eq.2}$$

$$\text{CRF}_i = \frac{d(1+d)^{n_i}}{(1+d)^{n_i} - 1} \quad \text{Eq.3}$$

$$\text{EAC} \left[\frac{\text{€}}{\text{MWh}}\right] = \frac{\text{OpEx}_{\text{yearly}} + \text{CRF}_n \cdot \sum_i \text{CapEx}_i \cdot \frac{n}{t_{d,i}}}{\text{E}_{\text{yearly}}} \quad \text{Eq.4}$$

$$\text{CRF}_n = \frac{d(1+d)^n}{(1+d)^n - 1} \quad \text{Eq.5}$$

with $E_{yearly}$ the yearly energy transported, $d$ the applied discount rate, $n_i$ the asset-specific depreciation time and $n$ the assessment period.

An alternative to the EAC approach is based on the net present value principle. This method is utilized by reputable commercial commentators such as Lazard[©] and Ernst and Young[©] [86,92]. While it involves a more complicated data setup, it offers more flexibility for the inclusion of inflation and complex cost assignment schemes. In contrast to the standard EAC method, this method, by default, requires the choice of a well-defined assessment period ($n$), but can more conveniently integrate equipment replacements and residual values. All asset acquisitions and replacements are booked at the correct year during the project, leading to a more intuitive accounting scheme. However, transparency about the inclusion of residual value at the end of the project remains essential. Finally, the time resolution of cash flows may offer better insight into the investment timeline and time discounting can be done more intuitively.

$$\text{LCOT}_{\text{NPV}} \left[\frac{\text{€}}{\text{MWh}}\right] = \frac{\sum_{t=0}^n \frac{\text{CapEx}_t + \text{OpEx}_t}{(1+d)^t}}{\sum_{t=0}^n \frac{E_t}{(1+d)^t}} \quad \text{Eq.6}$$

Under certain specified conditions, the EAC (Eq.2) and LCOT$_{\text{NPV}}$ (Eq.6) method are equivalent [92]. In section A2.1, we provide mathematical proof of this claim in the context of a techno-economic assessment of systems comprising multiple assets with different depreciation times. Equation 2 and 5 become mathematically equivalent when (1) the depreciation period for each asset in the EAC method is equal to the assessment period in the LCOT$_{\text{NPV}}$ method and no salvage value or decommissioning costs are considered, (2) yearly energy generation (or transport) remains constant over the assessment and (3) capital investments are done in year 0. Especially assumption (1) is unlikely to be met in assessments of hydrogen supply chains. Consequently, to confidently compare results from different studies, in our opinion, attention must be paid to the sensitivity of the results towards the assessment method. To date, this concern has been overlooked in sensitivity analyses of hydrogen supply chain techno-economic assessments.

*2.1.2.2   Default methodology*

By default, the LCOT is calculated using the EAC method (Eq.2) in order to align our method with the majority of the available literature. Additionally, we included the option to use the adapted EAC (Eq.4 and 5) and $LCOT_{NPV}$ approach, providing the user with the opportunity to experiment with and evaluate the effect of different costing methods on the estimated LCOT. Table 3 summarizes the general economic assumptions in this study. 2022 was chosen as the reference year of this study. The discount rate choice has a large impact on the cost outcomes of renewable energy projects assessments [95,96]. In the context of techno-economic assessments of private investments, the discount rate is often interpreted as the minimal acceptable return for that investment, including the opportunity cost of invested capital and possible risk premiums [97]. However, in the context of a public investment by a government, one often applies the social discount rate (typically 3-5%), representing a broader societally optimal rate at which cash flows should be discounted in projects with social relevance or that are funded through public channels. The social discount rate is low in comparison to corporate discount rates in order to safeguard benefits in the near and more distant future and avoid preference towards projects with high initial gains and persistent losses for future generations. The appropriate choice of discount rate thus depends on the perspective from which the assessment is made. In the default methodology of the assessment in this work, we assume a commercial perspective using region-specific weighted average cost of capital (WACC) as discount rate [98]. Large energy importing countries are often economically developed countries with a relatively low cost of capital. For energy exporters, generalization is not appropriate. Both highly developed economies like Australia or Oman, and emerging economies like Chile or Namibia, are determined to supply the worldwide hydrogen market [20,99–101]. As the intention of this work is to provide a more generalized insight into hydrogen supply chain costs, without considering specified routes, we chose a default general discount rate of 8% and provide the opportunity to alter this value in the application dashboard. Finally, it is not the intention of this work to evaluate business models. In corporate accounting, depreciation is used to allocate the costs incurred for an asset over its useful lifetime. Typically, accelerated depreciation schemes like the Modified Accelerated Cost Recovery System (MACRS) are used to write off a larger portion of the costs in the first years after the acquisition of the asset in order to lower the tax burden during the first years of operation [102]. The scope of this study does not include corporate income evaluation. Instead, the study aims to provide robust insight into the underlying cost dynamics with respect to hydrogen supply chain input variables. Therefore, we use the depreciation concept to reflect the economic reality of the asset (taking into account time preference), rather than as a tool for tax optimization or accounting. To this purpose, we have included the possibility to apply a linear depreciation scheme running for the assets' estimated useful lifes (for the adapted EAC and LCOT$_{NPV}$ method, the standard EAC method depreciates assets over the whole useful lifetime). The choice for a linear depreciation scheme is motivated by the assumption that the supply chains provide a constant transportation volume or linear value accumulation throughout the project lifetime and that the infrastructure can be used beyond the span of the assessment period if its lifetime allows it. When the assets lifetime is shorter than the project duration, replacement of the asset is incorporated. Based on the asset depreciations, residual values are booked as negative costs (revenues).

**Table 4:** General economic and accounting assumptions.

| Reference year | 2022 |
|---|---|
| Discount rate | 8% |
| Project duration | 20 years |
| Depreciation | linear |

Each process capital investment (CI$_0$) from reference data is adapted to the reference year and supply chain scale according to Eq.7. The relative scale of the reference supply chain (Sc$_0$), with respect to the scale of our model (Sc$_1$) is taken into account considering a scaling factor (SF). Finally, the cost estimation is adapted to include inflation effects through the Chemical Engineering Plant Cost Index (CEPCI) [103]. Because of geopolitical tension and (post)pandemic supply chain woes, the reference year 2022 was characterized by particularly high inflation. This translated in a dramatic spike in the CEPCI, a factor that needs to be taken into account when evaluating the CapEx cost contributors.

$$CI = CI_0 \left(\frac{Sc_1}{Sc_0}\right)^{SF} \left(\frac{CEPCI_{2022}}{CEPCI_0}\right) \quad \text{Eq.7}$$

The OpEx cost contribution is calculated as the sum of the energy requirements for the (de)conditioning steps, the energy consumption of storage after conditioning (for instance during transport), and general OpEx contributions like fuel consumption and labour costs. A general overview of the cost contributions and process steps that are incorporated into the model is provided in Table 5.

Table 5: General overview of the costs contributions incorporated in the different transportation supply chains.

| Marine transport | | GH$_2$ | LH$_2$ | LOHC | Ammonia | HCH |
|---|---|---|---|---|---|---|
| **Production facility** | *Conditioning* | Compression | Compression | Compression | Haber-Bosch process | Compression |
| | | - | Cooling | Hydrogenation | - | Cooling |
| | | - | Liquefaction | - | - | - |
| | *OpEx* | Maintenance | Maintenance | Maintenance | Maintenance | Maintenance |
| | | Insurance | Insurance | Insurance | Insurance | Insurance |
| | *CapEx* | Buffer tanks | Buffer tanks | Buffer tanks | Buffer tanks | Buffer tanks |
| | | Compression installation | Liquefaction installation | Hydrogenation plant | Haber-Bosch plant | Compression installation |
| | | - | Cooling installation | - | - | Cooling installation |
| | | - | Hydrogen pumps | - | - | - |
| | *Losses* | - | Boil-off | Deconditioning heat | Deconditioning heat | - |
| **Transport** | *OpEx* | Fuel | Fuel | Fuel | Fuel | Fuel |
| | | Labor | Labor | Labor | Labor | Labor |
| | | Maintenance | Maintenance | Maintenance | Maintenance | Maintenance |
| | | Insurance | Insurance | Insurance | Insurance | Insurance |
| | *CapEx* | GH$_2$ ship | LH$_2$ ship | Tanker ship | Ammonia ship | HCH ship |
| | | Pressure tanks | - | - | - | Clathrate tanks |
| | *Transport agent* | - | - | LOHC | - | Hydrophobic material |
| **Consumer** | *Deconditioning* | - | - | Dehydrogenation | Ammonia cracking | - |
| | *OpEx* | Maintenance | Maintenance | Maintenance | Maintenance | Maintenance |
| | | Insurance | Insurance | Insurance | Insurance | Insurance |
| | *CapEx* | Buffer tanks | Buffer tanks | Buffer tanks | Buffer tanks | Buffer tanks |
| | | - | Regasification plant | Dehydrogenation plant | Cracking facility | Compression installation |

## 2.1.3 Energy consumption (de)conditioning

The specific energy requirement of the different (de)condition processes included in this base case model demonstration is shown in Figure 7. The energy requirements are reported on the electricity level, after factoring in electrolysis at 52.59 kWh$_e$/kgH$_2$ efficiency. The graph shows that LOHC technology requires the most energy to perform (de)conditioning of the transported hydrogen. LOHC dehydrogenation is the main contributor for energy use in the LOHC chain, amounting to ca. 16 kWh$_e$/kg when hydrogen burning is considered. It is often assumed in the literature that the dehydrogenation energy demand can significantly be lowered when large amounts of waste heat are available at the dehydrogenation plant [18,54]. However, given the considered process temperature of 573 K it is unlikely that large amounts of waste heat can drive the endothermic reaction. Consequently, reference data including dehydrogenation powered completely by waste heat was not considered. Details on the reference data used for the LOHC chain energy requirements can be found in Table A.4. Ammonia (de)conditioning energy use estimates are still a subject of active debate in the literature. Especially energy use estimates for the cracking step vary substantially. Based on 7 independent literature sources (Table A.5), the energy need for ammonia cracking is estimated to be 10.07 kWh$_e$/kg. This value includes pressure swing adsorption purification and subsequent compression to 50 bar. For the LH$_2$ storage process, combining 10 reference sources (Table A.3), an energy use of 8.34 kWh$_e$/kg is incorporated. This leaves room for improvement, as current liquefaction plants achieve energy efficiencies between 10-13 kWh$_e$/kg [36,38]. The GH$_2$ process, requiring only compression and compressor cooling, is energetically much more favorable. Finally, the conceptual HCH technology also has a particularly low energy requirement for the (de)conditioning.

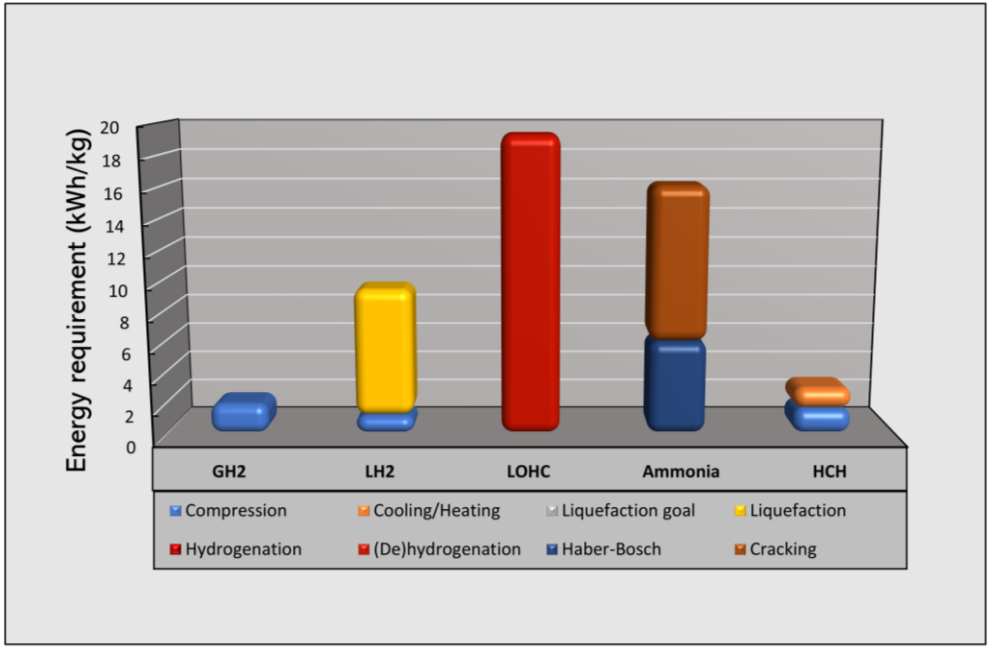

**Figure 7:** Specific energy consumption of the different hydrogen (de)conditioning technologies in the marine case.

## 3   Results & discussion

### 3.1   Supply chain costs

Figure 8 shows the LCOT of fossil-powered marine supply chains transporting hydrogen over 10 000 km at a 1 GW$_e$ capacity. At long overseas distances, the energy density of the carriers is the dominant cost determining parameter. As a result of low gravimetric energy density, GH$_2$ ship transport capacity is low and a large fleet is built in order to reach the supply chain's transportation capacity. Next to the large scale of the fleet, the unit cost of GH$_2$ ships is high owing largely to steel costs associated with the high-pressure vessels (see Table A.1). LH$_2$ supply chain costs are dominated by CapEx for liquefaction infrastructure and buffer storage. The need for buffer storage is inversely correlated with the capacity of the transportation ships, as increasing transport capacity lowers the docking frequency of ships and thus increases the need for storage capacity storing production *ad interim*. While only a low number of ships is required, the high unit cost of LH$_2$ carriers results in considerable ship CapEx cost. Given the low energy pices assumed in the producer's region, extra hydrogen production to account for boil-off losses does not lead to a high cost contribution. Ammonia has the lowest LCOT for the standard case in Figure 8. By far the largest cost contribution originates from Haber-Bosch and cracking infrastructure, followed by (de)conditioning energy use. Its high energy density (17.6 wt%) results in a low number of required ships and associated fleet costs. LOHC technology needs a larger transportation fleet as the energy density is lower (6.2 wt%). However, the unit cost for LOHC ships is also lower due to the favorable diesel-like characteristics of the dibenzyltoluene (DBT) carrier molecule and the scaled-up availability of suitable transportation ships. The cost difference with ammonia is largely caused by the cost for the storage agent, i.e., DBT. Analogous to the $GH_2$ case, HCH technology is more costly because of its large transportation fleet. The operation of the large fleet results in elevated fuel and labor costs.

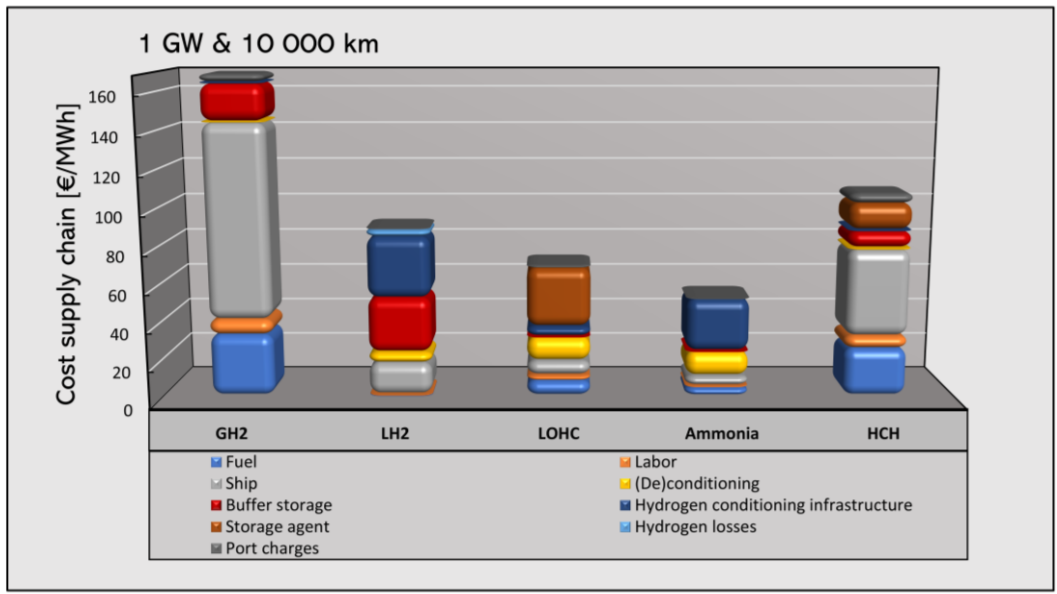

**Figure 8:** Cost breakdown of marine supply chains for a transportation distance of 10 000 km.

## 3.2 Scenario and sensitivity analyses

Hydrogen supply chains are complex systems dependent on multiple unit operations affecting the total transportation cost. In the previous section, we discussed the results for the default settings of the model. While Figure 8 revealed clear cost drivers for each technology, the presented cost breakdown is not generally applicable. This work aims to provide generalized insight into the cost dynamics of hydrogen supply chains. Generalized insight into cost dynamics can be achieved by identifying key cost drivers for each technology through rigorous scenario and sensitivity analyses. Exploring relevant supply chain scenarios, we found that the LCOT, and the respective cost drivers per technology behind the LCOT, are strongly affected by a wide space of supply chain input variables. In earlier studies, transportation distance, supply chain scale, storage agent costs and CapEx assumptions were frequently identified as sensitive parameters [10,12,18,27]. However, the number of investigated parameters and their input range variation is often very limited. Additionally, the impact of parameters like fuel choice and assessment method have thus far been overlooked.

Depending on the research field, definitions and procedures for uncertainty, sensitivity and scenario analysis may conflict. Following standard sensitivity analysis practice, many of the consulted studies on hydrogen supply chains have performed sensitivity analysis by varying input parameters by 20% or less in both directions [10,17,18,20,34]. Given the degree of uncertainty in input parameters like CapEx data (Figure 1), we believe sensitivity analysis over a wider range of reasonable input data would more intuitively convey the impact that can be established from the wide span of reasonable input parameters. Additionally, current studies on hydrogen supply chains mostly perform single parameter sensitivity analyses and do not appreciate interaction effects. Next to multiparameter sensitivity analysis, we believe extensive scenario analysis is needed to assess possible supply chain scenarios.

Without the constraints of a highly specific use case, the large amount of possible supply chain iterations for the 5 different technologies prohibits a satisfactory sensitivity analysis in the context of this text. To enable exploration of the multiparameter sensitivity of conceptual future hydrogen supply chains, the reader is provided with a user-friendly model interface allowing for detailed 'DIY' scenario analyses, building upon the modelling framework developed with the literature data referenced in this work. By means of example, the following sections provide in-depth analysis of a selection of parameters, deemed interesting by the authors. Figure 9 provides a qualitative overview of technology sensitivities found for a range of key input parameters. All these parameters can be manipulated in the interactive tool.

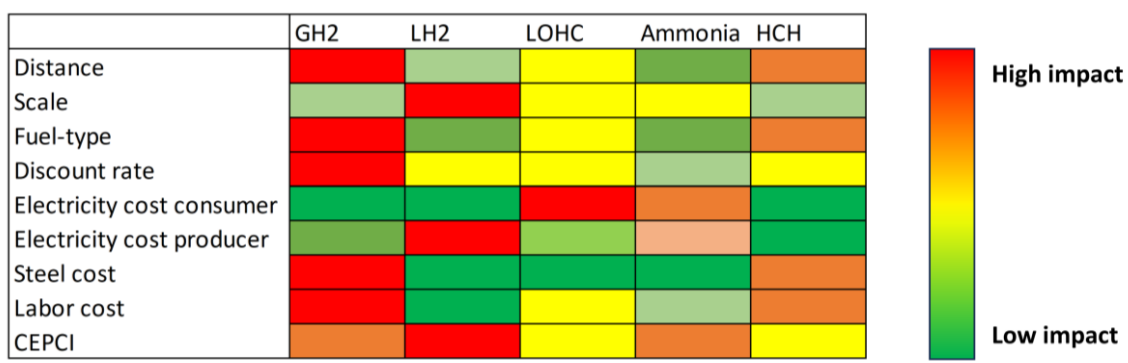

**Figure 9:** Heat map summarizing qualitative sensitivities of the supply chain costs of the different technologies towards prominent parameters relevant in case and local conditions setting.

### 3.2.1 Supply chain distance and scale

An increase in supply chain transportation distance, is typically associated with a rise in delivery time and lesser availability of transportation vehicles. Consequently, more vehicles must be deployed, thus causing a surge in LCOT. The need for fleet scaling is directly proportional to the energy density of the transportation technology, while the capital intensity of fleet expansion differs substantially between different technologies. Presently, hundreds of hydrogen supply chain routes are considered in the context of a worldwide hydrogen economy, ranging from a few hundred km to 25 000 km. Consequently, distance is typically considered as one of the most sensitive modelling parameters in hydrogen supply chain assessments. Also supply chain scale initiates fleet scaling, in a similar manner as supply chain distance. Contrary to distance, however, supply chain scale also directly affects stationary infrastructure costs.

The cost dependency of the marine supply chains on distance and supply chain scale is shown in Figure 10. In marine transport, the cargo capacity of a ship is indicated by weight or dead weight tonnage (DWT), since strict regulations are in place concerning the depth at which a cargo ship is allowed to sit in the water. Hence, transport using technologies having low gravimetric energy density (container included) can be limited by weight regulations, which will result in the need for quick scaling of the fleet. This effect can be witnessed in the graphs presented in Figure 8. $GH_2$ and HCH graphs contain many steps, with each step expressing the need for expansion of the fleet by one ship. Because the clathrate ships need considerably less steel for the pressure vessels, the cost per ship is lower for the HCH transportation chains. Additionally, because the pressure vessels are more lightweight, the clathrate ships can carry more pressure tanks. In order to facilitate discussion of the trends in Figure 10a, three distance ranges are identified: short distance (1 000-9 000 km), intermediate distance (9 000-17 000 km) and long distance (17 000 km-25 000 km). Each distance range is assigned a distinct color for clarity. For short distances up to 2 000 km, HCH and LOHC are the most cost effective. $GH_2$ quickly becomes the most expensive option as a result of the limited capacity of the ships. At distances larger than 2 500 km, ammonia becomes the most cost effective technology. The exceptional energy density of ammonia, combined with relatively low-cost ships, results in an almost flat cost curve. Figure 10b shows the effect of supply chain scale (input energy). $LH_2$ technology is the most prominent benefactor of scale increase at a fixed distance of 2 000 km (Figure 10b). At scales above 2 GW, scale effects lower $LH_2$ CapEx costs sufficiently for entering competitive cost ranges. HCH is the most cost competitive technology at scales below 1 GW. At intermediate (10 000 km) and long distances (19 000 km) (Figure 10b), ammonia is the lowest-cost technology for all scales. Especially at long distances, the LCOT of ammonia is uncontested at all scales.

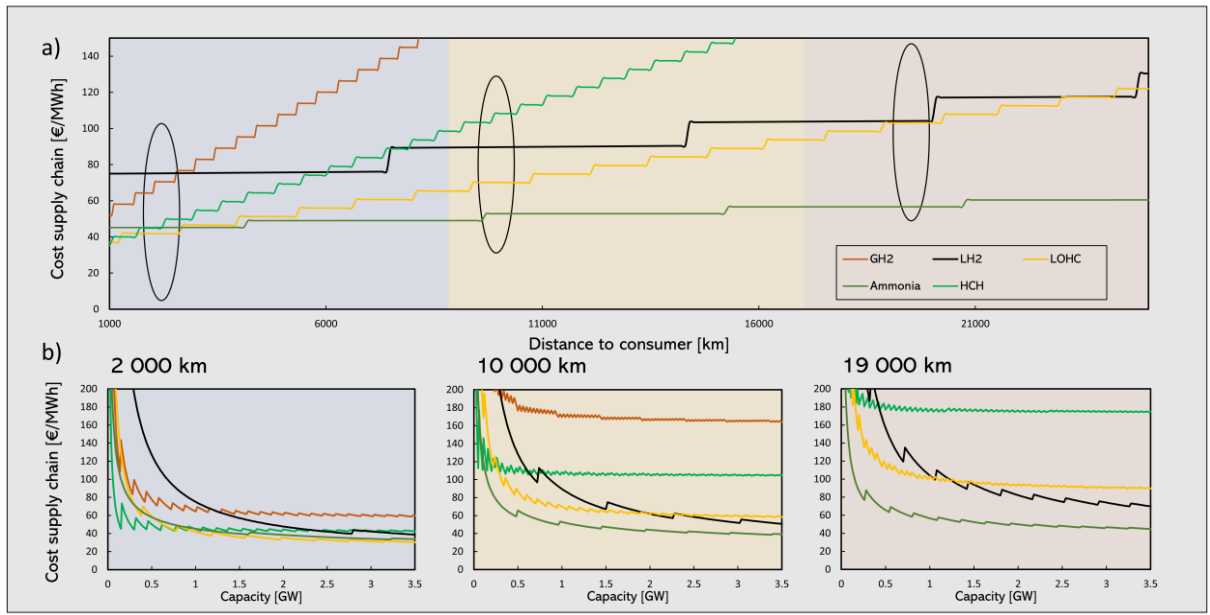

**Figure 10:** Distance dependency of marine transportation supply chain cost per MWh delivered hydrogen energy.

### 3.2.2 Impact cost assessment method

In section 2.1.2.1, the conditions for equivalence between the EAC and $LCOT_{NPV}$ method are introduced. When no residual values or decommissioning costs are considered, one of the conditions dictates that the depreciation period of each asset in the EAC method must be equal to the assessment period of the $LCOT_{NPV}$ method. In the standard EAC method, this condition can only be met if the useful lifetime for each asset in the supply chain is equal to the assessment period of the $LCOT_{NPV}$ method. In hydrogen supply chains, however, the involved unit operations can have different useful lifetimes and depreciation times. Expensive large-scale chemical infrastructure like Haber-Bosch plants or thermal crackers can have operational lifetimes exceeding 40 years. Their lifetime is often the result of project durations or within longer projects, careful cost-benefit analysis weighing increased maintenance and modernization costs versus renewal of installations. In the default settings of this work, it is assumed that the chemical (de)conditioning infrastructure, i.e., the liquefaction plant, LOHC (de)hydrogenation reactors, and ammonia synthesis and cracking infrastructure, is replaced after 20 years, while the assessment period is conveniently set equal to those operational lifetimes. As a result, the condition for equivalence of assessment methods is met in default settings of the EAC, EAC2 and $LCOT_{NPV}$ for the ammonia supply chain. In most real scenarios, however, this equivalence will not hold. Figure 11 introduces situations where the useful lifetime exceeds the assessment period. To include residual value at the end of the assessment period, a linear depreciation scheme is used in the EAC2 and $LCOT_{NPV}$ method. For the standard EAC method the useful lifetime of the plants is increased. The graph shows the effect of increasing depreciation periods on the LCOT of the ammonia supply chain. When the infrastructure is depreciated in 20 years, the condition for equivalence of assessment methods is met and all methods are equivalent. As depreciation periods increase, supply chain costs lower. However, the decrease in LCOT is much stronger for the EAC2 method. This discrepancy is caused by the fact that, in the LCOT$_{NPV}$ method, the remaining value of the asset at the end of the assessment period is booked in the final year of the assessment and is discounted accordingly. In the EAC2 method, time preference of cash flows is not directly considered. For the standard EAC method, no residual values can be booked, as there is no specified assessment period. However, the LCOT lowers because costs are spread over longer useful lifetimes.

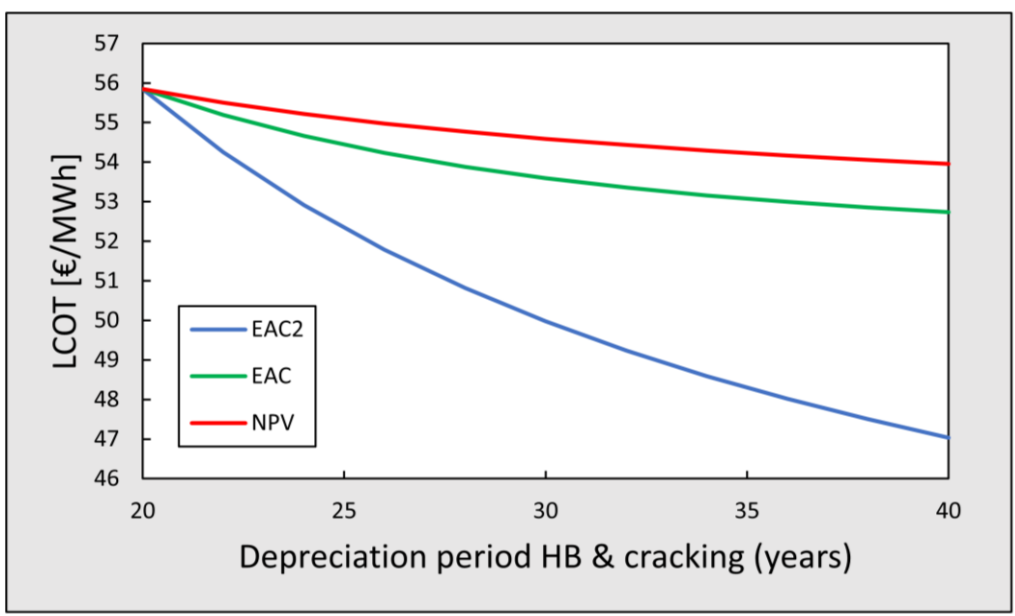

**Figure 11:** Effect of changing depreciation time of ammonia (de)conditioning infrastructure on ammonia supply chain cost for three different techno-economic assessment methods.

Depending on model settings, the techno-economic assessment method can thus have an important impact on the modelling results. When looking at current case studies on hydrogen supply chains in the literature, it is too often insufficiently clear if and how infrastructure replacements and depreciation is handled. Given its implications on the final results, as highlighted in Figure 11, dissemination of techno-economic analysis results (for academic or policy purposes) should always be accompanied by transparent documentation on and a clear motivation of the assessment method used.

### 3.2.3 Interaction effects

*3.2.3.1 Energy cost producer region*

Figure 8 shows a very low-cost contribution for the cost attribute "hydrogen losses" in the total cost of the $LH_2$ supply chain, which means that in the default scenario of this study, hydrogen boil-off from the ships and buffer storage can be compensated for at low cost. Figure 12 depicts the sensitivity of the "hydrogen losses" cost contribution to changes in buffer tank boil-off rate and producer region energy cost. Boil-off rates can vary depending on uncontrollable variables such as climate and they are strongly dependent on technology performance. Figure A.2 summarizes 11 reference sources for $LH_2$ stationary boil-off in boxplot format. While dedicated renewable electricity generation capacity is often planned for green hydrogen production and long-term Power Purchase Agreements might lower the risk for strong energy cost perturbations in the producer region, geopolitical tensions and local energy shortage may still result in large swings in energy cost over the project duration. Performing a single-parameter sensitivity analysis (white lines in Figure 12) in default settings of the model does not reveal strong sensitivity towards either parameter. However, one can clearly appreciate the interaction effect of both parameters when they are increased simultaneously, with the hydrogen loss cost escalating to extreme values within the considered parameter ranges. At an energy cost of 0.10 €/kWh (high-end LCOE off-shore wind [104]) and a boil-off rate of 0.5% per day (maximal non-outlier in Figure A.2), hydrogen losses amount to 30 €/MWh (point A in Figure 12), which represents a 1000% increase with respect to the default "hydrogen losses" cost contribution, and a 36% increase with respect to default total $LH_2$ supply chain costs. The reason behind this strong interaction effect is that boil-off losses are compensated for by extra production in the producer region. Considering the high energy costs in the consumer region, it is far more cost effective to compensate for boil-off during transportation or at the consumer facility by foreseeing extra production and transport, rather than reliquefying boil-off on site.

While this section might not reveal a surprising synergy, the extent to which costs can escalate merits prudence. The example in this section is provided to illustrate the complexity of interacting cost effects in techno-economic assessments with many degrees of freedom. Given a certain modelling framework, single parameter sensitivity analyses are sometimes insufficient to provide confidence in the resilience of modelling results under input parameters like energy cost or technology performance, that can vary substantially within cases, and more particularly between different case studies. Comprehensive scenario analysis or multi-parameter sensitivity analysis protocols like the Sobol method might therefore be required to realistically assess the impact of volatile input parameters on complex techno-economic models [105].

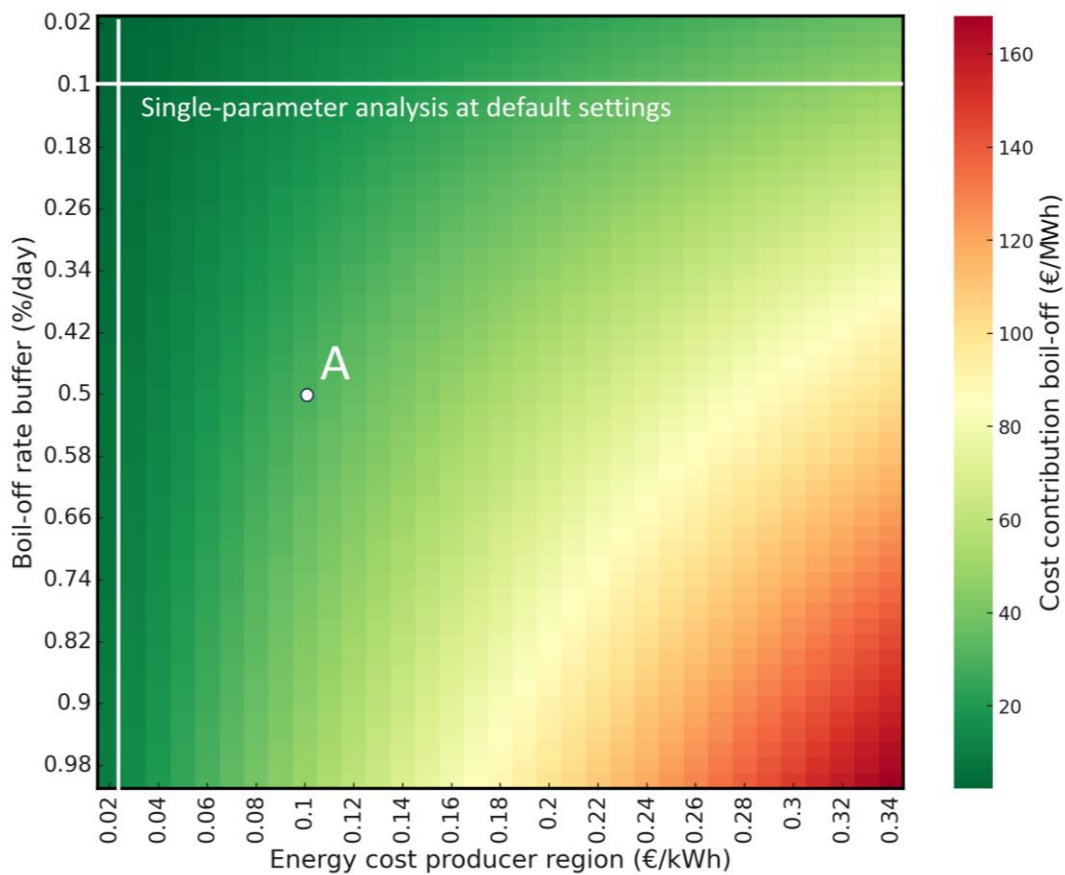

**Figure 12:** Heatmaps showing the sensitivity of the cost contribution "Hydrogen Losses" for LH$_2$ technology with respect to buffer boil-off rate and producer region energy cost. The white lines represent single parameter sensitivity analyses for both parameters, keeping the other one at default model settings.

*3.2.3.2  Impact of fuel transition*

Another interesting interaction effect that can be identified is the impact of a transition to clean fuels on the cost evolution of the different technologies with increasing supply chain distance. In the following years, the clean energy transition will rapidly change energy system needs. Western industrialized economies could need sizeable imports of clean molecules. Timely establishment of supply chains would be needed to avoid delays to the energy transition in such a scenario. Therefore, the assessment of hydrogen supply chains must be seen in the right timeframe. Fossil fuels for marine transportation will not be banned in the coming years and they must be considered to correctly assess competitivity of technologies at the moment of commissioning of the first applications. To date, most techno-economic analyses therefore assume fossil-powered propulsion systems (Table 1). Figure 13 compares the cost evolution with distance for supply chains powered by renewable fuels with their fossil-fuel-powered alternatives. A first observation that can be made is that the cost evolution of all fossil-powered technologies is nearly linear, while renewably powered supply chains show an increasing slope throughout the distance range. This effect is caused by the assumption that the renewable drivetrains use their own cargo as fuel. With increasing distance, less deliverable capacity remains and thus costs increase more sharply. Importantly, these scaling discrepancies are more articulate for the low-density carriers. Assuming all ships will eventually be powered by renewable fuels, assessing fossil powered drivetrains can thus lead to a significant bias towards low energy density technologies. As ammonia fuel's volumetric energy density approaches fossil fuels more closely, little effect is witnessed in ammonia supply chains. Since using hydrogen boil-off is the most economical choice in both scenarios, no distinction is made for the $LH_2$ supply chains.

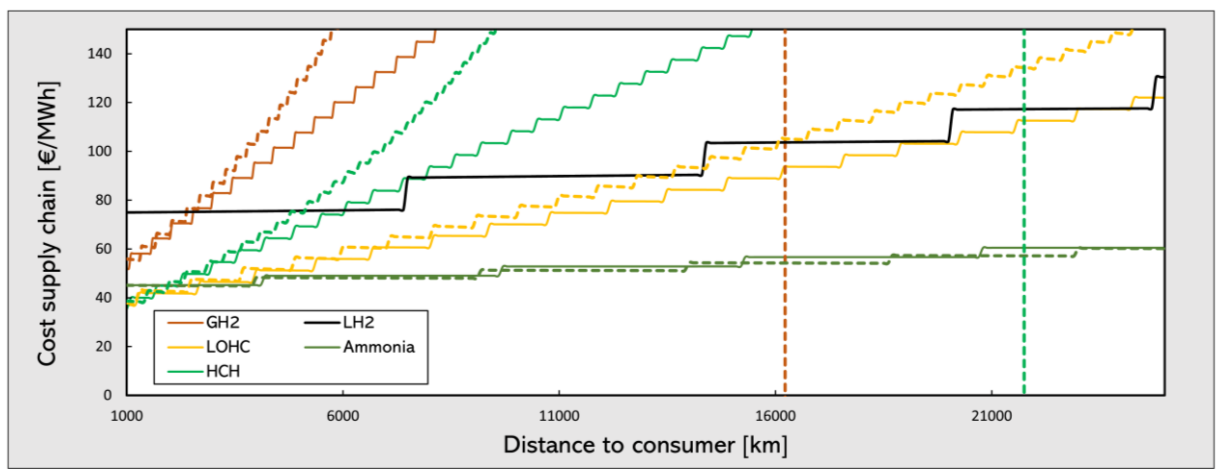

**Figure 13:** Comparison between supply chain cost evolution of fossil-powered marine supply chains (solid lines) with respect to their renewably powered alternatives (dashed lines). The vertical dashed lines indicate the maximal distance $GH_2$ and HCH ships can travel using their own cargo as fuel.

## 3.3 Limitations and future work

The present work brings perspective to the techno-economic literature on marine hydrogen supply chains by critically highlighting challenges and limitations of existing studies and approaches. Simultaneously also the present framework comes with its own limitations. For example, it only covers the handling and delivery of hydrogen after it has been produced. The production itself is not included in the model (except for an estimation of costs linked to compensation of hydrogen losses in the different supply chains). When comparing specific routes using the presented framework, it is therefore harder to incorporate local conditions for hydrogen production. Practically, longer, more expensive routes might be lucrative if the costs associated with local hydrogen production are lower. In future versions of the toolbox, a production module could be incorporated.

Another limitation of the toolbox concerns result validation. The prevalence of input uncertainty and modelling variability and connected thereto, output variability renders benchmarking of our results extremely difficult. One-on-one comparison of our specific results will only yield accidental quantitative confirmation. To maximize our modelling robustness, each development step of the code included elaborate coding logic checks, looking into edge cases and comparing outcomes with common sense and the outcome range found in literature.

The current model architecture also lacks optimization for automated execution of large-scale sensitivity and scenario analyses, which are essential for protocols like Monte Carlo or variance-based sensitivity assessments. Such assessments can increase the robustness of the results and identify more interaction effects. Improved automation scripts could enable such analyses in future versions. The web-app will also be updated when new data becomes available, and at some point also alternative energy carriers like HVDC lines and hydrogen pipelines should be implemented. Finally, the present model does not provide information on the sustainability of the different supply chains. In order to improve the value of the outputs for policymakers, it would be interesting to add a Life Cycle Assessment functionality to assess the environmental impact of the alternatives.

# 4 Conclusion

Hydrogen is considered a key energy vector that could assist decarbonizing the global energy system. To optimally plan the integration of hydrogen energy in the global energy system, policymakers, industry and energy system researchers require, unbiased and widely applicable insight on energy transport cost dynamics. While in recent years, many studies have modelled hydrogen supply chains, looking for the most cost effective transportation technologies and supply chain configurations for specific trading routes, the vast span of possible supply chain scenarios and configurations, in combination with the cost uncertainty related thereto, complicate generalized conclusion-making. Interestingly, even within similar case studies, supply chain cost outcomes diverge strongly. Most of the variability is caused by the many degrees of freedom in the design of a hydrogen supply chain assessment. Variability can come from case and model specification choice, local condition volatility or uncertainty in technology-related data. The large variation in technology performance estimations is often caused by differing (1) views on technology progress, (2) operational input/output parameters (temperature, pressure, hydrogen purity,…), (3) modelling detail, and (4) investor perspective (commercial, public). Additionally, information on these factors is not always available. As a result, it is hard to make comprehensive conclusions on supply chain competitivity across the different available studies. Building on this premise, a meta-analysis model was developed to allow dynamic investigation of hydrogen supply chains under varying input scenarios. By providing a convenient model-interface that leverages an increasing amount of underlying cost data, users are enabled to identify and gain insight into more generalized cost dynamics attributes of hydrogen transportation technologies. Authors of new studies are encouraged to send in cost data and scenarios to be included in the cost database.

The model is demonstrated in default settings, for which technology-specific characteristics and cost drivers are identified and discussed by performing a selection of sensitivity analyses deemed interesting by the authors. Under the range of conditions studied, some general technology characteristics are noteworthy. $GH_2$ technology is fit only for short-distance supply chains, as costs escalate quickly due to its low gravimetric energy density. While stationary infrastructure cost is low, costs are dominated by CapEx for storage tanks. Due to its high energy efficiency, however, $GH_2$ can be an option for short distance routes sourcing from regions with relatively high energy cost. $LH_2$ technology cost is dominated by liquefaction plant and storage costs, and scale effects strongly increase $LH_2$'s cost competitivity. Additionally, low energy cost in the producer region is important as hydrogen liquefaction is by far the most energy intensive unit operation in the supply chain. LOHC (DBT) technology is favored by low fleet expansion cost per ship and convenience, as current diesel carriers can be used. However, the mediocre energy density leads to sub-optimal scaling at the fleet level. Ammonia carrier ships are both relatively low-cost and energy dense, which leads to cost-effective scaling. Supply chain costs are dominated by stationary infrastructure cost. Finally, HCH technology shows strong similarities with $GH_2$ technology, but is favored by more cost-efficient scaling due to improved gravimetric energy density.

In techno-economic analyses, the goal is to translate technical complexities in given macro-economic conditions into a non-biased cost indicator, which can be compared to technology alternatives. Multiple methods are available to distil the same cost structures into an overall cost indicator. The two most popular methods used in hydrogen supply chain analyses were incorporated into the presented model and conditions for mathematical equivalence were investigated. Analyses outside these conditions showed that the different time discounting strategies in cost assessment methods can cause wide cost divergence when depreciation time of expensive infrastructure is not matched with the assessment period of the analysis. Clear communication on depreciation and time discounting strategies should therefore always be included in techno-economic assessments involving cost-intensive infrastructure with particularly short or long lifetimes.

Figure 14 shows the cost evolution of the lowest-cost technology of the model in default settings throughout the considered distance range, combined with the optimal technology choice at specified distances from studies considered in this work (Table 1). It is clear that there is no consensus in the literature when correcting for distance. The LCOT of our findings in default settings varies between 30-66 €/MWh (roughly 1-2 €/kgH$_2$), without considering hydrogen production cost. While physical bounds limit energy efficiency of hydrogen supply chains indefinitely, large cost improvements for the required infrastructure can be achieved with scale and maturity in the hydrogen supply chain market. The sustainable energy and chemical feedstock transition will require a vast span of solutions. It is the task of academics to provide policymakers with unbiased information and tools to make informed and future-proof decisions for each part of the global transition towards climate neutrality. Whether, and to what extent, hydrogen imports are a part of that transition will depend on geopolitics, economics and capital deployment across dispatchable energy alternatives and chemical feedstock options. With this work, we want to provide insight into hydrogen transportation economics, based on the currently available literature. The interactive application dashboard is available to gain insight into the vast span of possible scenarios involved in hydrogen supply chain economics and can be freely accessed online [106].

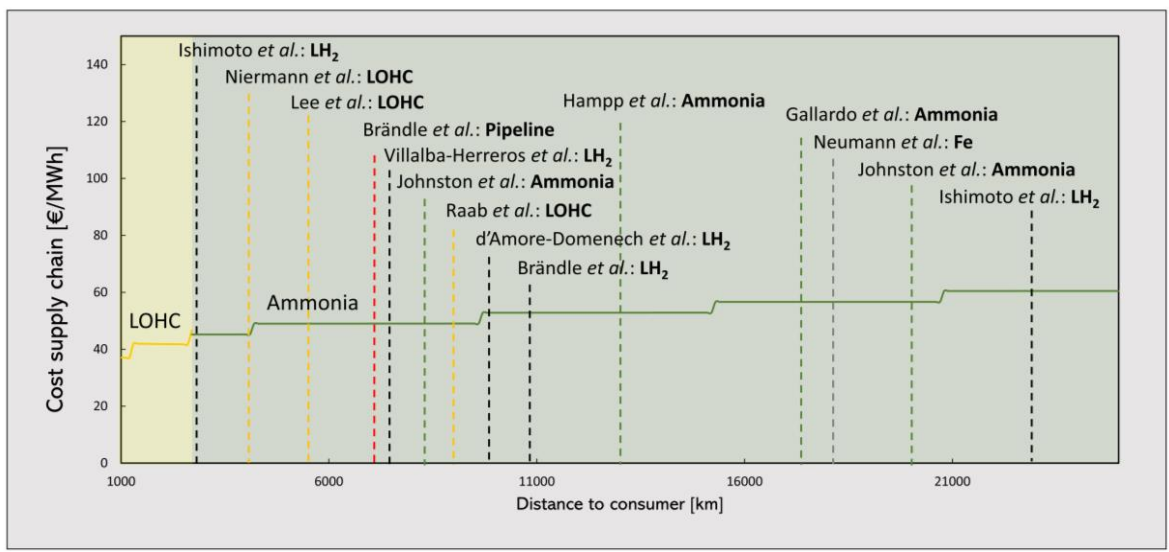

**Figure 14:** Evolution of best case LCOT for the default settings of the model, combined with best technology outcomes of case studies in literature. The position of the literature sources on the graph is not indicative of the quantitative LCOT in the studies, as different scope and input parameters render quantitative comparison uninformative.

## Supplemental information

Document A1. Figures A.1-A.4. Tables A.1-A.14.

## Acknowledgement

The authors acknowledge VLAIO for Moonshot funding (ARCLATH, n° HBC.2019.0110, ARCLATH2, n° HBC.2021.0254). M.H. acknowledges FWO for an FWO-SB fellowship. J.A.M. acknowledges the Flemish Government for long-term structural funding (Methusalem) and department EWI for infrastructure investment via the Hermes Fund (AH.2016.134). NMRCoRe acknowledges the Flemish government, department EWI for financial support as International Research Infrastructure (I001321N: Nuclear Magnetic Resonance Spectroscopy Platform for Molecular Water Research). J.A.M. acknowledges the European Research Council (ERC) for an Advanced Research Grant under the European Union's Horizon 2020 research and innovation programme under grant agreement No. 834134 (WATUSO).

## Credit author statement

L.Hanssens: Conceptualization, Investigation, Methodology, Formal analysis, Validation, Original Draft and Visualization. K. Van Acker & M.Houlleberghs: Validation and Review & Editing. L. Hanssens & E. Breynaert: Implementation of the web-application. E.Breynaert & J.A.Martens: Conceptualization, Validation, Writing – Review & Editing, Supervision, Funding acquisition.

## Declaration of competing interest

The authors declare that the research was conducted in the absence of any commercial, financial or personal relationships that could be construed as a potential conflict of interest.